# Landslide vulnerability analysis using frequency ratio (FR) model: a study on Bandarban district, Bangladesh.

Nafis Fuad[a], Javed Meandad[b], Ashraful Haque[c], Rukhsar Sultana[d], Sumaiya Binte Anwar[e], Sharmin Sultana[f]

[a] PhD Fellow, Civil and Environmental Engineering, Wayne State University

[b] Landslide Early Warning System (LEWS) Coordinator, Caritas Bangladesh

[c] Coordinator, FOREWARN Bangladesh

[d] Manager Monitoring, MIS, Project Support and Reporting, BRAC

[e] Program Manager at Center for Participatory Research and Development (CPRD)

[f] Jatiya Kabi Kazi Nazrul Islam University

*Corresponding author: nafis_fuad@wayne.edu

**Landslide vulnerability analysis using frequency ratio (FR) model: a study on Bandarban and Rangamati districts, Bangladesh.**

Abstract

This study assesses landslide vulnerability in the Chittagong Hill Tracts (CHT), specifically focusing on Bandarban district in Southeast Bangladesh. By employing a multidisciplinary approach, thirteen factors influencing landslides were examined, including terrain features, land use, and environmental variables. Utilizing the FR model and integrating various datasets such as DEM, satellite images, and rainfall data, landslide susceptibility mapping was conducted. The analysis revealed that steep slopes, high elevations, specific aspects, and curvature contribute significantly to landslide susceptibility. Factors like erosion, soil saturation, drainage density, and human activities were also identified as key contributors. The study underscored the impact of land use changes and highlighted the stabilizing effect of vegetation cover. The resulting Landslide Susceptibility Map (LSM) categorized the area into five susceptibility zones. The model demonstrated a prediction accuracy of 76.47%, indicating its effectiveness in forecasting landslide occurrences. Additionally, the study identified significant changes in the study area over three decades, emphasizing the influence of human activities on slope instability. These findings offer valuable insights for policymakers and land-use planners, emphasizing the importance of proactive measures to mitigate landslide risks and ensure community safety. Incorporating these insights into policy frameworks can enhance the effectiveness of mitigation strategies in the vulnerable regions.

## Introduction

Landslides, among the most impactful natural hazards globally, pose significant threats to both human lives and socioeconomic development. With approximately 9% of global disasters attributed to landslides, their devastating effects are particularly prevalent in mountainous regions (Yilmaz 2010; Kanwal et al. 2017). Landslides represent a significant geological hazard with substantial implications for the socioeconomic advancement of numerous countries (Lazzari et al. 2018). This geological process encompasses a diverse array of ground movements, including ground creep, rockfall, translational and rotational failures (slides), rock topples, and debris/earth flows, which can transpire under various geological and topographical settings (Cruden and Varnes 1996). Typically, the predominant force driving landslides is gravity; however, multiple additional factors influence the initial slope condition. The occurrence of landslides is widely acknowledged across both developing and developed regions worldwide (Chen et al. 2018).

Bangladesh, a country highly susceptible to diverse natural hazards, faces considerable challenges posed by landslides, especially in its hilly areas, such as the Chittagong Hill Tracts (CHT) (Rabby et al. 2020). Located in the southeastern part of the country, CHT comprises three districts: Rangamati, Khagrachari, and Bandarban (Rabby and Li 2019a). These hilly regions, with elevations ranging from 600 to 900 meters above sea level, experience multiple landslides each year, leading to loss of life, property damage, and economic losses (Sultana 2020). Intense precipitation during the monsoon serves as the primary precipitating event for landslides in the region (Khan et al. 2012). The global climate crisis has aggravated the severe and prolonged rainfall, resulting in landslides occurring with escalated frequency in recent periods (Abedin et al. 2018; Ahmed et al. 2020). Additionally, the role of individuals in inducing hillslope instability is unparalleled, as evidenced by housing expansion on the hillslope, hill excavation, heightened population pressure, deforestation, unsustainable agricultural practices, urban expansion, and deficient governance (Rahman et al. 2017).

Landslide is the movement of rock, soil, and debris downslope under the influence of gravity. It is a natural phenomenon and depends on various factors, including local geology, topography, climate, and land use/land cover type (Rabby and Li 2019b). Prolonged rainfall and earthquakes are the primary triggers of landslides. Road construction on the slopes, hill cutting, and deforestation are the major anthropogenic activities that create a conducive condition for landslides (Rabby and Li 2019b). It is a global phenomenon that plays a significant role in the landscape's evolution. However, in many areas, they also pose a severe threat to localized people directly, mainly influenced by different phenomena (Alexander 2005). Recent studies have ranked landslide as the 4th biggest deadly among natural disasters, after floods, storms, and earthquakes (Sultana 2020).

In recent years, landslides in the CHT have been on the rise, posing a severe threat to the local population. The increasing trend of frequency and damage is mainly triggered by much higher rainfall amounts than the monthly average, particularly during the monsoon season (Ahmed 2015). The geological formation and soil characteristics of the hilly areas in CHT, comprising yellowish-brown to reddish-brown loams composed of unconsolidated sedimentary rocks, make these regions highly susceptible to landslides (Sultana 2020). Anthropogenic activities, including housing expansion, hill excavation, population pressure, deforestation, and unsustainable agricultural practices, further contribute to hillslope instability (Rahman et al. 2017).

The areas have a complex soil composition. The hilly soils are mainly yellowish-brown to reddish-brown loams that are composed of unconsolidated sedimentary rocks such as sandstone, siltstone, shale, and conglomerate with local unconformities that make these regions extremely susceptible to a landslide (Sultana 2020). The physical, social, political, and environmental settings of this area are diverse. The Chattogram Hill Tracts area is a place of 11 tribes that are utterly dependent on the hilly environments for their livelihoods. Traditional shifting cultivation (locally known as Jhoom), conducted by the tribal communities, is also another leading cause for landslides as it declines the forest cover and enhances inappropriate land use that leads to severe soil erosion (Sultana 2020).

Given the complex terrain and its implications for human settlements and livelihoods, there is an urgent need to understand and assess landslide vulnerability in the Bandarban and Rangamati districts of Bangladesh. Landslide susceptibility mapping is a crucial tool for proper land use planning and disaster management in the region (Guzzetti et al. 1999). Over the years, various techniques and methods have been developed and applied in the literature to map landslide susceptibility. Among these, the frequency ratio (FR) model, a probabilistic approach, has gained popularity due to its effectiveness in combination with Geographic Information Systems (GIS) and Remote Sensing (RS) techniques (Lee and Sambath 2006). This model has been commonly used to determine landslide susceptibility zones based on landslide inventory data and relevant conditioning factors, such as local geology, topography, climate, and land use/land cover type (Bui et al. 2013; Zêzere et al. 2008).

In this context, this study aims to conduct a comprehensive landslide vulnerability analysis in the Bandarban and Rangamati districts, Bangladesh, using the frequency ratio (FR) model. By integrating landslide inventory data with relevant environmental and anthropogenic factors, the study seeks to identify and map areas with varying degrees of landslide susceptibility. The findings of this research are expected to provide valuable insights for disaster preparedness, land use planning, and sustainable development in the region, ultimately contributing to the reduction of landslide-induced risks and enhancing the resilience of local communities.

There is a lack of strict hill management policy within CHT regarding forest cleaning, tree plantation and implementation of engineering measures to protect demolished hills, regulating the growth of residential areas including other construction activities (Sultana, 2020). This has encouraged many informal settlements along the landslide-prone hill-slopes in Chittagong. Study by Sultana (2020) shows that more than 500,000 impoverished people are living in informal settlements on the risky foothills of Chattogram (Khan 2008; Islam 2018b). People are living on the hillslopes and hilltops illegally by overlooking the landslide risks as most of them are landless and poor (Sultana, 2020). These settlements are being considered as illegal by the formal authorities, while the settlers claim themselves as legal occupants. This kind of conflict has also weakened the institutional arrangement for reducing the landslide vulnerability in Chittagong City. An acute land tenure conflict has been ongoing among the public agencies, settlers, powerful elites and the local community representatives over the past few decades.

Mapping the areas that are susceptible to landslides is essential for proper land use planning and disaster management for a locality or region. Throughout the years, different techniques and methods have been developed and applied in the literature for landslide susceptibility mapping. Landslide susceptibility maps can be produced using either quantitative or qualitative approaches. There are mainly four methods available to map landslide susceptibility, namely **landslide inventory based (i) probabilistic**, (ii) **deterministic**, (iii) **heuristic**, and (iv) **statistical techniques** (Guzzetti et al. 1999) of these techniques, the probabilistic and statistical methods have been commonly used in recent years. These methods have become more popular, assisted by Geographic Information Systems (GIS) and Remote Sensing (RS) techniques (Lee and Sambath 2006). Probabilistic models like frequency ratio (FR), bivariate analysis, multivariate analysis, and Poisson probability model (Bui et al. 2013) are more frequently used to determine the landslide susceptibility zones (Zêzere et al. 2008). Afterwards the most vulnerable map was generated through overlaying the landslide susceptible map and the composite index maps.

The significance of this study lies in its potential to inform policymakers and local authorities about the areas most vulnerable to landslides, enabling them to develop targeted measures for disaster preparedness, risk reduction, and urban planning. By addressing the critical issue of landslide vulnerability in Bandarban and Rangamati districts, this research endeavors to contribute to the overall resilience and sustainable development of the Chittagong Hill Tracts region in Bangladesh.

**Methodology**

**Study area**

The Chittagong Hill Tracts (CHT), located in the Southeast part of Bangladesh, is known for its very steep, rugged, and mountainous terrain (UNICEF, 2019). The region comprises three hill districts: Rangamati, Bandarban, and Khagrachari. For the purpose of this study, two study locations within Rangamati and Bandarban were chosen to assess the landslide vulnerability of CHT. These two cities, Rangamati and Bandarban, have already been recognized as highly vulnerable to landslides, with several devastating incidents impacting city dwellers. Since 2017, landslides have resulted in the loss of nearly 219 lives in various informal settlements within Rangamati and Bandarban city and adjacent small urban centers (Prothom Alo).

Rangamati, covering an area of 6116.13 sq km, is situated between 22°27' and 23°44' northern latitudes, and 91°56' and 92°33' east longitudes. It shares borders with the Tripura state of India to the north, Bandarban district to the south, Mizoram State of India, and China Pradesh of Myanmar to the east, and Khagrachari and Chittagong districts to the west. A 77 km road connects Rangamati to Chittagong (Hossain and Hossain 2018). The district Administration of Rangamati is divided into 10 Upazilas (sub-district), 49 unions (sub-sub district), 162 mauzas, 1555 villages, 2 paurashava, 18 wards, and 90 mahallas (Bangladesh Bureau of Statistics, 2015). For this study, the focus lies on three Upazilas: Kawkhali, Naniarchar, and Kaptai, which were selected for landslide vulnerability analysis.

Bandarban District, covering an area of 4479.03 sq. km, is located between 21°11' and 22°22' northern latitudes, and 92°04' and 92°41' east longitudes. It shares borders with Rangamati district to the north, Arakan (Myanmar) to the south, Chin Province (Myanmar), and Rangamati district to the east, and Chittagong and Cox's Bazar district to the west (Hossain and Hossain 2018). The district comprises 7 upazilas, 30 unions, 96 populated mauzas, 1554 villages, 2 pourashava with 18 wards, and 102 mahallas (Bangladesh Bureau of Statistics, 2015). The selected upazilas for the assessment are Bandarban Sadar, Lama, Rowangchhari, Ruma, and Thanchi.

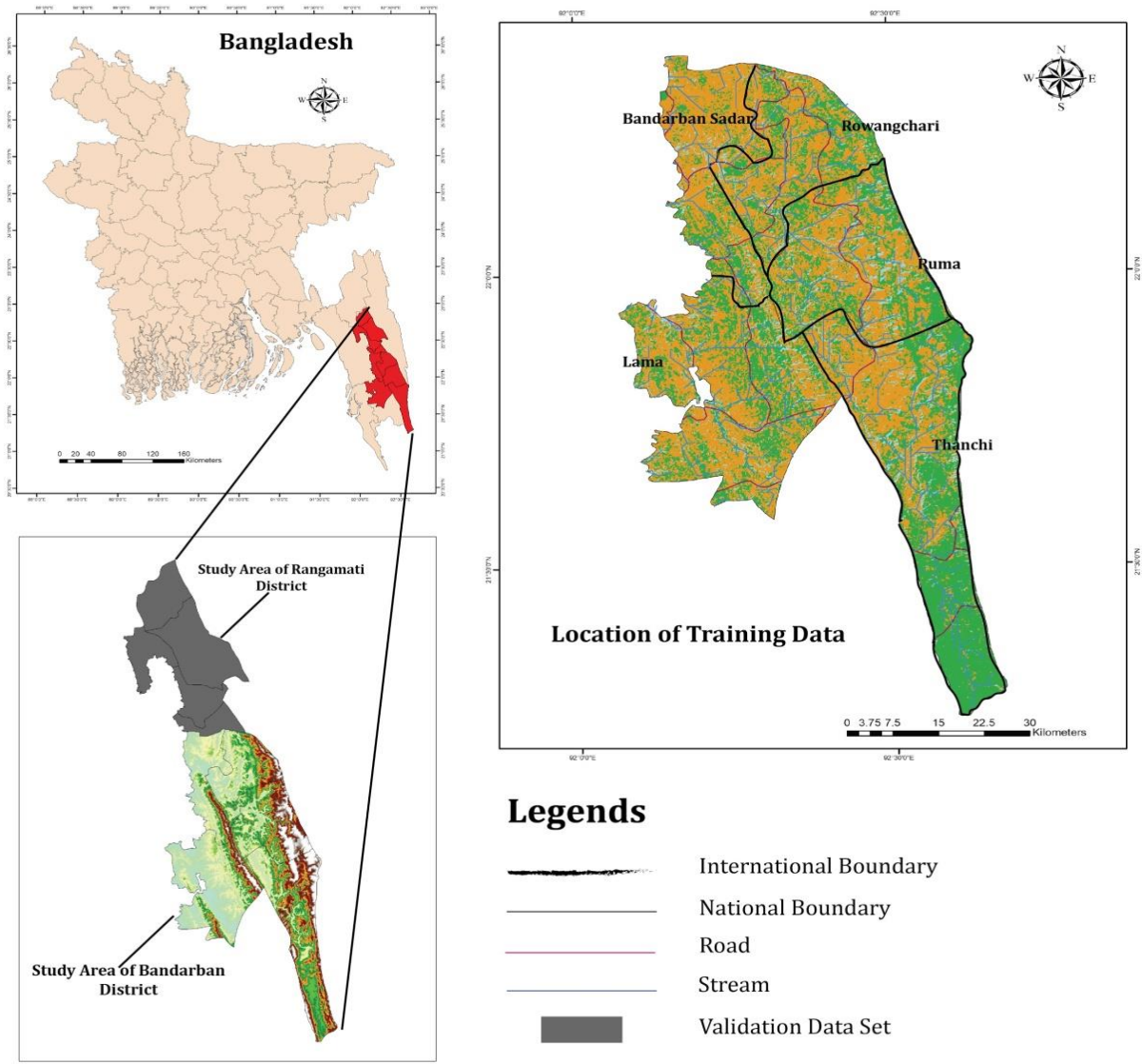

**Fig.1** Location of the study area

**Data and methods**

The present work was accomplished by carefully applying various secondary data sources. To validate the results, a certain portion of the study area was considered for the training data set. Global Positioning System (GPS) was also used to identify landslide locations in the study area. Historical landslide records and other relevant information were obtained from secondary sources and PRA exercise.

Shuttle Radar Topography Mission (SRTM) digital elevation model (DEM) data with a resolution of 30 m * 30 m was obtained from the United States Geological Survey (USGS) to prepare elevation, slope aspect, slope curvature, slope angle, Stream Power Index (SPI), and Topographic Wetness Index (TWI) layers. Geological data were obtained from the Geological Survey of Bangladesh (GSB), and soil properties data were acquired from the Soil Resource Development Institution (SRDI). Landsat TM Images from different years (1989, 2000, 2010, and 2020) were used to prepare land use maps, NDVI maps, and cut & fill maps. Rainfall data of the monsoon period (1998-2010) were extracted from the 2B31 layer of Tropical Rainfall Measuring Mission (TRMM) satellite imagery. Additionally, other data such as road networks, administrative boundaries, and market locations were extracted from secondary sources.

To derive the frequency ratio for each class, all the data layers were incorporated with the prepared landslide inventory map. All the data layers were projected in an ARC GIS environment. The study utilized the Bangladesh Transverse Mercator (BTM) projection system with Python suitability scripts. The methodologies for each parameter as stated above have been provided in detail in the following sections (Table 1).

**Table 1. Database of the current study**

| Datasets | Parameters | Source | Scale or resolution | Classification method |
|---|---|---|---|---|
| Shuttle Radar Topography Mission (SRTM) and digital elevation model (DEM) | Slope | USGS Earth Explorer | 30 m × 30 m | Natural break |
| | Elevation | | 30 m × 30 m | Natural break |
| | Aspect | | 30 m × 30 m | Equal interval |
| | Curvature | | 30 m × 30 m | Natural break |
| | Stream Power Index (SPI) | | 30 m × 30 m | Natural break |
| | Topographic Wetted Index (TWI) | | 30 m × 30 m | Natural break |
| | Drainage distance | | 30 m × 30 m | Natural break |
| | Drainage density | | 30 m × 30 m | Natural break |
| | Euclidean distance of road | | 30 m × 30 m | Natural break |
| Rainfall data | Rainfall | 2B31 layer of Tropical Rainfall Measuring Mission (TRMM) satellite imagery | 30 m × 30 m | Natural break |
| Landsat TM Images | Land Use | USGS Earth Explorer | 30 m × 30 m | Supervised classification |
| | NDVI | | 30 m × 30 m | Supervised classification |
| | Cut fill | | 30 m × 30 m | Supervised classification |

**Landslide Conditioning Factors**

In this research, we carefully selected fifteen conditioning factors, considering existing literature, their effectiveness, data availability, and relevance to landslide occurrence. These factors encompass elevation, slope aspect, slope angle, slope curvature, geology, soil properties, distance to road, cut & fill, stream power index (SPI), Stream distance, stream density, topographic wetted index (TWI), monsoon rainfall, normalized differential vegetation index (NDVI) and land use. By incorporating all these chosen factors, we conducted landslide susceptibility mapping.

To ensure uniformity in data analysis, each conditioning factor was converted into a raster format with a spatial resolution of 30 × 30 m. Subsequently, we classified the data using the Jenks natural breaks method within the ArcGIS application. This approach helps in identifying meaningful groupings and patterns without any bias.

**Slope angle**

Slope angle is a crucial factor in landslide occurrences (Dai and Lee 2002; Gomez and Kavzoglu 2005). Landslides happen when gravitational force overcomes resisting force (Islam et al. 2017). Slope stability depends on shear strength, determined by cohesion, internal friction, and material strength (Wang et al. 2016; Du et al. 2017; Islam 2018a). The steepest stable slope is defined as the angle of repose (Islam 2018a). Slopes between 20 to 60 degrees are more susceptible to sliding, while angles of 70 degrees or greater are common in landslide-prone areas like Chittagong (Islam 2018a). Landslide occurrences concentrate at gully heads and steep overhanging landforms (Bajracharya and Maharjan 2018). Changes in slope stability can be caused by various factors, acting together or alone. Steep hills are vulnerable to sliding. Hills with slopes of 20 to 30° have moderate susceptibility, while those with slopes of 40 to 60° have high susceptibility; Chittagong hills have slopes of 70° or greater (Islam 2018b) (Fig. 2). Most landslides occur at the gully head of the catchment and on steep overhanging landforms (Bajracharya and

Maharjan 2018). As slope angle increases, the likelihood of landslides also rises (Xu et al. 2012; Pham et al. 2017; Panchal and Shrivastava 2021).

**Elevation**

Elevation is another crucial factor influencing landslide susceptibility mapping, as it affects various environmental conditions on slopes, including human activities, vegetation, soil moisture, and climate dynamics (He et al. 2012; Dou et al. 2015; Shu et al. 2021). The study area's elevation was classified into five categories, ranging from 7 to 1021 m. The extreme northwestern and western parts are dominated by higher elevations, ranging from 501 to 1021 m (Fig. 2). In the southernmost section, the Bay of Bengal and its surrounding areas in the western part, there are high elevations from Myanmar hills, with extensive high vegetation coverage. Moderate elevation zones are found in the mid and southern parts of the study area. The largest area was covered by low elevation zones, followed by moderate and higher elevations.

**Curvature**

The surface curvature of the study area, derived from DEM data, plays a significant role in shaping the topography and influencing the hydrological conditions of the soil (Oh et al. 2017). Based on the classification shown in Fig. 2, the curvature was categorized into three types: concave slopes (negative values), convex slopes (positive values), and flat surfaces (Saleem et al. 2019). In this context, positive values signify upwardly convex slopes at specific cells, while negative values indicate upwardly concave slopes. A zero value denotes a linear surface (Mersha and Meten 2020). Notably, concave slopes possess the ability to retain water more effectively and have a higher potential for landslides compared to convex slopes (Lee and Choi 2004; Mersha and Meten 2020).

**Slope aspect**

Slope aspect describes the direction of slope in an area and plays a crucial role in landslides and exposure to sunlight, drying winds, rainfall, and discontinuities (Du et al. 2017; Khan et al. 2019; Shu et al. 2021). In previous research, a link was found between slope aspect and its proneness to landslides (Alkhasawneh et al. 2013; Dou et al. 2015). Some landslide cases identified slope aspect as one of the main contributing factors for the occurrences of landslides (Alkhasawneh et al. 2013). The study area experienced several slope aspects, including flat-north, northeast-east, southeast-south, southwest-west, and northwest-north facing slopes (Fig. 2). Each slope had individual geomorphic characteristics in terms of soil erosion, run-off, drainage, sediment transport, slope angle, etc. All these geomorphic parameters were influenced by micro-climatic variations throughout the region. South, southwest, and southeast facing slopes received a significant amount of orographic rainfall, leading to drainage branching all over the slope and promoting surface run-off and soil erosion processes. Additionally, the disintegration and decomposition of the slope were caused by prevalent diurnal maximum solar radiation across all slope facets.

**Stream power index (SPI)**

Stream power index values were associated with factors such as water mass conservation, gravity, basin hydrology, hydraulic geometry, shear stresses, climate, concavity, flood interval, fracture spacing in bedrock, and bedrock erodibility (Moore et al. 1991; Fonstad 2003). Slope played a significant role in determining the stream power index, leading to variations at the reach scale (Fonstad 2003). The stream power index (SPI) was derived from the DEM and classified into five zones (Fig. 2). SPI represented the measure of the erosive power of the overland flow, assuming that discharge was proportional to the catchment area, and it predicted net erosion in the flow acceleration and convergence zones, while net deposition occurred in the zones of decreasing flow velocity (Pourghasemi et al. 2014). The stream power index is calculated with the following equation.

$$SPI = Ln (As*tan\beta)$$

where Ln is the natural log, As is the flow accumulation, and tan $\beta$ is the slope.

**Topographic Wetness Index (TWI)**

The Topographic Wetness Index (TWI) is another crucial factor in landslide susceptibility. It expresses the saturation of slope materials, indicating areas with water accumulation, including seasonally and permanently waterlogged ground saturation (Chen and Yu 2011; Das and Lepcha 2019). It reveals the geomorphic complexity of landslide terrain, highlighting topographic highs (dry areas) and lows (wet areas) (Sörensen et al. 2006). Higher TWI values indicate a greater tendency for slope materials saturation. TWI considers slope, upslope contributing area, and rainfall at a specific location, representing the flow accumulation within a watershed (Wilson and Gallant 2000). The study area was classified into five topographic wetted index zones (Fig. 2). The Topographic Wetness Index is calculated with the following equation.

Topographic Wetness Index (TWI) = $Ln (As / \tan \beta)$

where Ln is the natural log, As is the flow accumulation, and $\tan \beta$ is the slope.

**Rainfall**

Precipitation and run-off were closely related to the sudden inundation of hill slopes and landslides. All the major landslide events occurred at much higher rainfall amounts compared to the monthly average. Excessive rainfall resulted in an increase in ground water pressure, destabilizing the slope of the hill. Mass soil wasting was caused by gravity, as gravity exerted a force downward proportional to the amount of mass. Saturation of the pore spaces increased the weight of the mass, intensifying the sheer force and hazard intensity. When sloped areas were completely saturated with water, landslides could occur. In the absence of mechanical root support, the soils started to run off. The monsoon rainfall ranged from 568 to 2100 mm. Based on the rainfall distribution, the study area was classified into five zones (Fig. 2).

**Drainage distance**

The concentration of drainage network and their engagement in the process of erosion and transportation make the mountain slope more vulnerable to landslide phenomena. In the present study, five different buffer categories were made through Euclidean distance i.e., 0 – 454 m, 454 – 940 m, 940 – 1457 m, 1457 – 2052 m, 2052 – 3000 m (Fig. 2).

**Drainage density**

Drainage density was the length of stream per unit area of a river basin. Landslides were prominent in areas where drainage density was high and the soil layer was too thin (Onda 1993). The drainage density varied from one place to another based on the natural drainage network in the study area. Higher drainage density indicated more drainage concentration and slope saturation over the space. The formula, given below, was used to calculate drainage density (Horton 1932).

$D_d = L\mu / A$

where $D_d$ is drainage density, $L\mu$ is the length of the stream and A is the total area. The grid method has been used to calculate the specific drainage density of the basin (Fig. 2).

**Distance to road**

The distance to roads is considered one of the most significant factors affecting landslide occurrences in hilly areas (Li et al. 2021). Road construction near hill slopes can alter the natural conditions of the area, influencing landslide susceptibility. Landslides were observed along the road side due to the lack of proper roadside drainage, leakage in existing drainage, and under-capacity drainage. Other failures occurred due to blockage of drainage, toe cutting of slopes, and oversaturated sediments. The road network served as a lifeline to the people and the country's economy (Dikshit et al. 2020). The proximity to the road factor was critical as road cuts were quite common in this region, and the risk decreased as the distance increased away from the road network. The distance to the road was calculated using the 'Euclidean Distance' technique, which provided the distance from each cell in the raster to the closest source (Sander et al. 2010). The Euclidean distance tools gave information based on Euclidean or straight-line distance, calculated from the center of the source cell to the center of each of the surrounding cells. The distance to the road of the study area is classified into five classes (Fig. 2).

**Normalized Difference Vegetation Index (NDVI)**

The Normalized Difference Vegetation Index (NDVI) is a standardized measure of greenness (relative biomass) and vegetation density over space. It is useful in assessing vegetation cover and land use patterns, which are often influential in landslide occurrences due to anthropogenic interference on hill slopes (Dou, et al., 2015). NDVI values range from -1.0 to 1.0, where negative values are associated with clouds, water, and snow, and values near zero represent rock and bare soil. Low values (0.1 and below) indicate barren areas, moderate values (0.2 to 0.3) represent shrub and grasslands, while high values (0.6 to 0.8) indicate temperate and tropical rainforests.

The NDVI was estimated using the formula

$NDVI = (IR − R)/(IR + R)$,

Here, IR is the infrared portion and R is the red portion of the electromagnetic spectrum. The positive values of NDVI indicate vegetation health whereas negative values represent water depth.

In the study area, NDVI ranges from −0.144 to 0.54 As illustrated in Fig.2. Low NDVI showed minimum vegetation cover and low vegetation density and maximum exposure of the surface to the atmospheric process.

**Land use**

Land use is another influential factor affecting landslide occurrences. We derived the land use map from Sentinel-2 satellite imagery using a supervised classification technique in ArcGIS, categorizing it into six classes (Fig. 2). The study area is predominantly covered with croplands and scrubs. In the last four decades, the Bangladesh government shifted a significant amount of the population from plain lands of different districts to hilly areas of CHT. The CHT region recorded a population growth rate of about

67.95% during these decades, while the national growth rate ranged from 3% to 6%. Among them, Bengali plain land people increased from 9% to 49% (Bangladesh bureau of statistics, 2015; Channel24, 2017a; Pertha, 2017a). Due to this increased population, there was a rapid increase in natural resources consumption for housing materials, which triggered deforestation in the area (Chakma and Chakma 2017). Besides, Jhooming, a traditional farming practice by the local tribal people, systematically destroyed the virgin forests in the study area. The exposed forest led to increasing vulnerability of loose soils to rainfall, eventually leading to landslides.

**Cut fill**

The cut-and-fill tool was used to identify changes in landform surface modification. The Cut Fill tool summarized the areas and volumes of change resulting from a cut-and-fill operation. By comparing surfaces of a given location at two different time periods, it identified regions of surface material removal, surface material addition, and areas where the surface had undergone changes. Landsat 4-5 TM DEM 1989 and Landsat 8 TM DEM 2020 were considered to find out land erosion and accretion in the study area (Fig. 2).

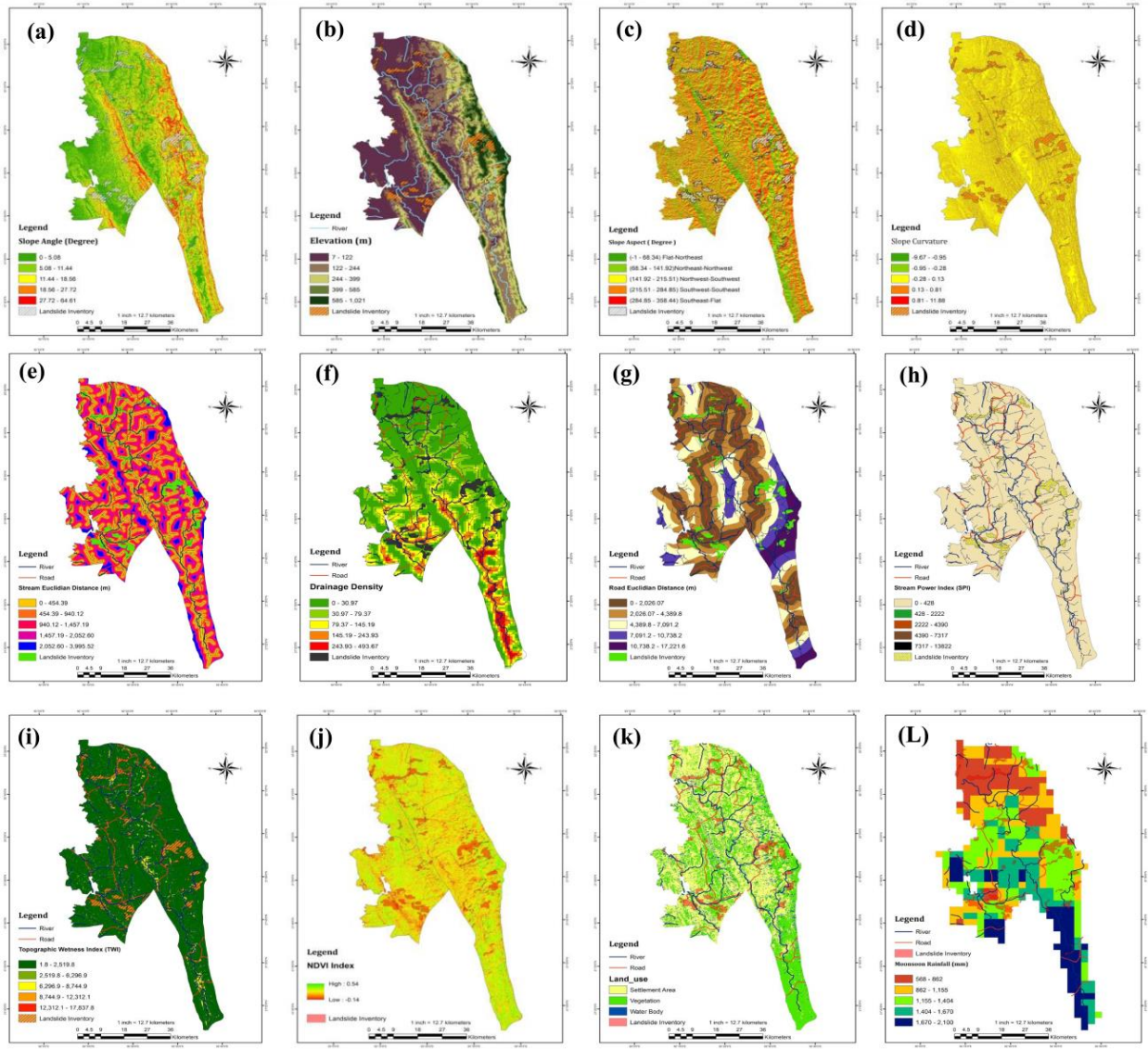

**Fig. 2.** Raster layers of relief factors (a) slope, (b) elevation, (c) aspect, (d) curvature, (e) stream euclidean distance, (f) drainage density, (g) road euclidean distance, (h) SPI, (i) TWI, (j) NDVI, (k) land use, and (L) rainfall

## Landslide Susceptibility Modeling

### Frequency Ratio (FR) Model

Frequency ratio (FR) is a highly popular method for assessing landslide susceptibility, widely used in natural hazard studies such as floods, landslides, and droughts (Li et al. 2017; Zhang et al. 2020; Sarkar et al. 2022). The FR model is a bivariate statistical analysis used for landslide susceptibility analysis, analyzing both landslide events and associated factors (Lee and Pradhan 2006). It involves analyzing the ratio of landslide-affected pixels to the total pixels in specific geographical units. To create a landslide susceptibility zonation map, each class of landslide conditioning factors was considered along with the associated landslide and non-landslide pixels (Lee and Talib 2005; Regmi et al. 2014). By establishing combinations between the landslide inventory map and criterion maps, the FR for each factor's class was determined, indicating the probability and significance of landslides in a given location. A higher FR value signifies a stronger relationship between the factor and landslides, while a value below 1 indicates a weaker association. A value of 1 represents an average relationship between landslides and the study area (Mandal et al. 2018; Addis 2023). The frequency ratio has been calculated using Eq. (1)

$$F_{r_i} = \frac{\{N_{pix}(S_i)/N_{pix}(N_i)\} \times 100}{\{\sum N_{pix}(S_i)/\sum N_{pix}(N_i)\} \times 100}$$

where N pix (Si) is the number of pixels containing landslide in each class (i) N pix (Ni) is the total number of pixels having class (i) in the whole basin, ΣN pix (Si) total number of pixels in each class, and ΣN pix (Ni) total number of pixels in the whole basin (Regmi et al. 2014).

After calculating the frequency ratio, all the raster map parameters of frequency ratio have been summed up to make landslide susceptibility index value (LSIV) using the following equation (Pradhan, 2010).

LSI = $Fr_1$ + $Fr_2$ + $Fr_3$+………………+$Fr_n$

where LSI is landslide susceptibility index value and $Fr_1$, $Fr_2$, $Fr_n$ is the frequency ratio of the raster data layers and n is the total number of factors for the study.

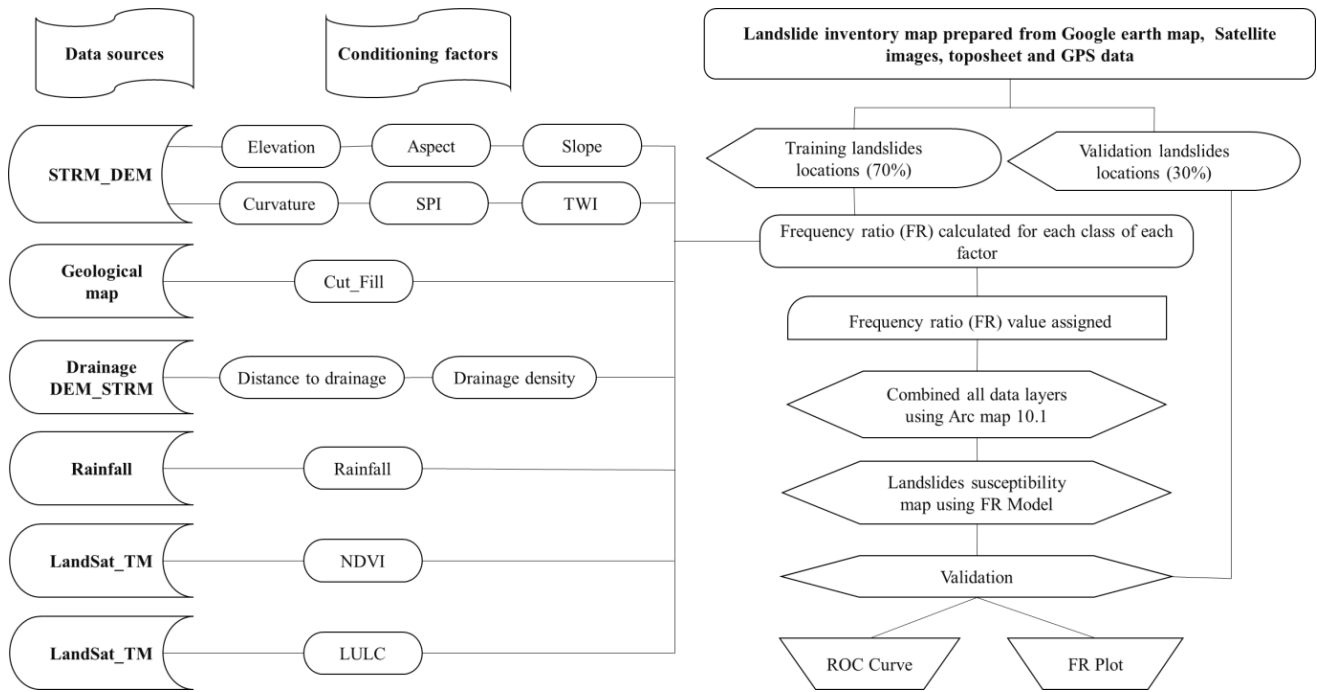

**Fig. 1** Methodology of frequency ratio (FR) model

**Models Validation**

The validation process of the model proposed in Fig. 2 was conducted by training the data set from the landslide inventory maps using the geostatistical tool. All the selected conditioning factors were assessed based on their Area under Curve (AUC). AUC is a type of accuracy statistics used for prediction models (probabilities) in the assessment and analysis of natural disaster events. The AUC value plays a significant role in determining which conditioning factors will be employed in landslide susceptibility mapping using the Frequency Ratio (FR) method. A higher AUC value indicates a greater level of statistical accuracy in the model. When the threshold definition reaches the maximum value of 1, it signifies optimal prediction accuracy, independent of the specific threshold chosen. In other words, a higher AUC value suggests a more reliable model in capturing landslide susceptibility through the FR method (Lepore et al. 2012; Mohammady et al. 2012; Rossi and Reichenbach 2016; Deng et al. 2017).

To assess the impact of conditioning parameters on landslide activity, a threshold-agnostic approach was adopted, employing the Receiver Operating Characteristic (ROC) curve. The ROC curve illustrates the precision values attained across various threshold values, and it was generated by utilizing validation and cross-validation datasets to approximate its area. The matrix values of each parameter were analyzed using SPSS statistics to derive the ROC curve (Pimiento 2010; Wahono 2010; Rossi and Reichenbach 2016). The calculated results were presented as a percentage of the study area classified as susceptible (x-axis) versus the cumulative percent of landslide occurrence (y-axis) using the AUC calculation formula as follows (Pimiento 2010).

$$AUC = \sum_{i=0}^{n}(x_i - x_{i-1})y_i - [(x_i - x_{i-1})(y_i - y_{i-1})/2]$$

where xi represents the percentage of the area, and yi represents the area of the landslide.

AUC is a graph of varying index numbers, usually ranging between a maximum value of 1 (or 100%) and 0.5 (or 50%). From the AUC values, the classification of the model can be determined, where 0.9 indicates a very good model classification, 0.8–0.9 indicates a good model classification, 0.7–0.8 indicates a medium or reasonable model classification, and < 0.6 indicates a poor model classification. Hence, conditioning parameters with a minimum AUC value of 0.6 are recommended. The higher the AUC value of a parameter, the greater its influence on the landslide event (Pimiento 2010; Silalahi et al. 2019).

## Results

### Frequency ratio model

The study results indicate several key relationships between different factors and landslide frequency ratio (FR) in the study area. Firstly, there is a positive relationship between slope and landslide FR, with slopes ranging from 0 to 11.44° and 27.72 to 64.61° having FR values greater than 1, indicating a high probability of landslide occurrences. Moderate slopes also showed FR values exceeding 0.95. Secondly, elevation played a significant role, with higher altitude areas having higher FR values, indicating a greater likelihood of landslides. The elevation class 585-1021m had the highest FR value (2.06), followed by 399-585m (1.59) and 244-399m (1.41), while 122-244m had the lowest FR value (0.71). Additionally, slope aspect and curvature were found to influence landslide susceptibility. Slopes facing northeast-east, southeast-south, southwest-west, and northwest-north exhibited FR values greater than 1, indicating heightened landslide probability. Negative curvature (concave) areas and positive curvature (convex) areas also showed high FR values, further contributing to landslide susceptibility.

Furthermore, there was a positive correlation between Stream Power Index (SPI) and landslide FR, indicating the study area's susceptibility to erosion and slope instability. Regarding drainage density, high FR values were found in certain classes, such as class 0 - 454.39 km/sq. km and 145.19 - 243.93 km/sq. km, indicating an increased probability of landslides in those areas. The study also revealed significant changes in the study area between 1989 and 2020, with only 10.25 percent remaining unchanged. Land surface cutting activities were prevalent in 53.03 percent of the area, while 36.708 percent experienced accretion. Both loss and gain areas exhibited higher FR values, highlighting their contribution to slope instability.

In the NDVI, the FR value is greater than one, where the NDVI classes -0.144 - 0.183, 0.183 - 0.261, and 0.261 - 0.321, indicating a high probabilities of landslides occurrence. However, the remaining NDVI classes have low FR value less than one; with relatively high vegetation coverage can easily lead to landslide occurrence. The relationship between TWI landslide probabilities showed that 2,519.8-6,296.9, 6,296.9-8,744.9 8,744.9-12,312.1, and 12,312.1-17,837.8 classes have the highest value of FR (1.43, 1.27, 1.81, and 1.14), r7espectively, greater than one. With regard to the conditioning factor rainfall, four classes with 568-862mm, 862 - 1,155mm, and 1,155 - 1,404mm have a higher FR value than the other classes and are the most landslide occurrence classes.

**Table: Class frequency ratios of selected parameters**

| Parameters | Classes | No of pixels $[N_{pix(N_i)}]$ | % of $N_{pix(N_i)}$ | Landslide pixels $[N_{pix(S_i)}]$ | % of $N_{pix(S_i)}$ | FR |
|---|---|---|---|---|---|---|
| Slope | 0 – 5.08 | 2548859 | 65.58 | 145249 | 65.7 | 1.00 |
| | 5.08- 11.44 | 1100810 | 28.32 | 63001 | 28.49 | 1.01 |
| | 11.44- 18.56 | 215571 | 5.54 | 11664 | 5.275 | 0.95 |
| | 18.56- 27.72 | 20050 | 0.51 | 1092 | 0.493 | 0.96 |
| | 27.72- 64.61 | 1236 | 0.03 | 73 | 0.033 | 1.04 |
| Elevation | 7 - 122 | 2506131 | 64.48 | 141193 | 63.865 | 0.99 |
| | 122 - 244 | 829189 | 21.33 | 33465 | 15.137 | 0.71 |
| | 244 - 399 | 379450 | 9.763 | 30408 | 13.754 | 1.41 |
| | 399 - 585 | 154881 | 3.985 | 14034 | 6.347 | 1.59 |
| | 585- 1,021 | 16878 | 0.434 | 1979 | 0.895 | 2.06 |
| Aspect | -69.34 | 687024 | 17.64 | 36492 | 16.48 | 0.93 |
| | 68.34 - 141.92 | 725869 | 18.64 | 42026 | 18.99 | 1.02 |
| | 141.92 -215.51 | 698690 | 17.94 | 40150 | 18.142 | 1.01 |
| | 215.51 -284.55 | 1040659 | 26.73 | 59760 | 27.003 | 1.01 |
| | 284.55 -358.44 | 740577 | 19.024 | 42875 | 19.373 | 1.02 |
| Curvature | -8.71 | 44 | 0.0011 | 4 | 0.0018 | 1.60 |
| | -0.67 | 79828 | 2.053 | 4487 | 2.0295 | 0.99 |
| | -0.41 | 3805633 | 97.918 | 216529 | 97.941 | 1.00 |
| | 0.13- 0.81 | 1016 | 0.026 | 58 | 0.02623 | 1.00 |
| | 0.81 - 11.88 | 6 | 0.00015 | 1 | 0.00045 | 2.93 |
| Stream Power Index (SPI) | 0 -428 | 2055576 | 52.88 | 118525 | 53.61 | 1.01 |
| | 428 - 2222 | 735465 | 18.92 | 40984 | 18.53 | 0.98 |
| | 2222 - 4390 | 400221 | 10.297 | 21919 | 9.91 | 0.96 |
| | 4390 – 7317 | 228237 | 5.872 | 12808 | 5.79 | 0.99 |
| | 7317 – 13822 | 467030 | 12.016 | 26843 | 12.141 | 1.01 |
| Topographic Wetted Index (TWI) | 1.84 - 2,519.8 | 3651684 | 93.95 | 202259 | 91.487 | 0.97 |
| | 2,519.8 - 6,296.9 | 162984 | 4.193 | 13270 | 6.0023 | 1.43 |
| | 6,296.9 - 8,744.9 | 56067 | 1.44 | 4043 | 1.828 | 1.27 |
| | 8,744.9 - 12,312.1 | 12751 | 0.328 | 1310 | 0.592 | 1.81 |
| | 12,312.1 - 17,837.8 | 3043 | 0.078 | 197 | 0.0891 | 1.14 |
| Drainage distance | 0 - 454.39 | 1108365 | 28.434 | 82703 | 37.34 | 1.31 |
| | 454.39 - 940.12 | 1050460 | 26.948 | 62803 | 28.35 | 1.05 |
| | 940.12 - 1,457.19 | 905290 | 23.224 | 41466 | 18.72 | 0.81 |
| | 1,472.19 - 2,052.60 | 600704 | 15.41 | 26170 | 11.81 | 0.77 |
| | 2,052.60 - 3,995.52 | 233206 | 5.982 | 8343 | 3.766 | 0.63 |
| Drainage density | 0 - 30.97 | 1856226 | 47.69 | 93810 | 42.35 | 0.89 |
| | 30.97 - 79.37 | 994850 | 25.56 | 59827 | 27.01 | 1.06 |
| | 79.37 - 145.19 | 603426 | 15.51 | 39177 | 17.68 | 1.14 |
| | 145.19 - 243.93 | 340732 | 8.75 | 26762 | 12.08 | 1.38 |
| | 243.93 - 493.67 | 96632 | 2.48 | 1909 | 0.861 | 0.35 |
| Euclidean distance of road | 0 - 2,026.07 | 1276117 | 32.73 | 72788 | 32.86 | 1.00 |
| | 2,026.07 - 4,389.83 | 1142515 | 29.31 | 71979 | 32.49 | 1.11 |
| | 4,389.83 - 7,091.27 | 797858 | 20.46 | 27129 | 12.24 | 0.60 |
| | 7,091.27 - 10,738.21 | 417186 | 10.71 | 25196 | 11.375 | 1.06 |
| | 10,738.21 - 17,221.67 | 264349 | 6.781 | 24393 | 11.013 | 1.62 |
| Land Use | Water Body | 291745 | 7.48 | 20372 | 9.197 | 1.23 |
| | Settlement Area | 2088060 | 53.56 | 131625 | 59.42 | 1.11 |
| | Vegetation | 1518220 | 38.94 | 69488 | 31.37 | 0.81 |
| Rainfall | 568 - 862 | 712204 | 18.42 | 41976 | 19.18 | 1.04 |
| | 862 - 1,155 | 891637 | 23.061 | 56904 | 26.014 | 1.13 |
| | 1,155 - 1,404 | 1009565 | 26.11 | 93087 | 42.55 | 1.63 |
| | 1,404 - 1,670 | 681406 | 17.62 | 20264 | 9.26 | 0.53 |
| | 1,670 - 2,087 | 571576 | 14.78 | 6509 | 2.97 | 0.20 |
| Cut fill | Net Gain | 1424648 | 36.708 | 85557 | 38.706 | 1.05 |
| | Unchanged | 398127 | 10.25 | 10185 | 4.607 | 0.45 |
| | Net Loss | 2058193 | 53.03 | 125301 | 56.68 | 1.07 |
| NDVI | -0.144 - 0.183 | 159981 | 4.1 | 11901 | 5.37 | 1.31 |
| | 0.183 - 0.261 | 555897 | 14.26 | 38034 | 17.17 | 1.20 |
| | 0.261 - 0.321 | 1137134 | 29.17 | 68588 | 30.96 | 1.06 |
| | 0.321 - 0.377 | 1271794 | 32.62 | 68171 | 30.77 | 0.94 |
| | 0.377 - 0.541 | 773218 | 19.83 | 34791 | 15.71 | 0.79 |

**Landslide susceptibility map (LSM)**

The final landslide susceptibility map was produced by combining the 13 landslide conditioning factors that are listed in Fig. The landslide vulnerability map is classified into five classes manually based on the probability of landslide occurrences on the field. In the landslide susceptibility map, 610667 pixels showed the very high susceptibility zone (16.49% area), 1007582 pixels high susceptibility zone (27.21% area), 1069399 pixels moderate susceptibility zone (28.88% area, 652020 pixels low (17.61% area), and 363867 pixels very low (9.82% area).

**Table:** Area under different landslide vulnerable classes based on frequency ratio model

| Level of susceptibility | Pixel | Area (sq. km) | Area in % |
|---|---|---|---|
| Very Low | 363867 | 327.48 | 9.82 |
| Low | 652020 | 586.818 | 17.61 |
| Moderate | 1069399 | 962.459 | 28.88 |
| High | 1007582 | 906.824 | 27.21 |

| Very High | 610667 | 549.6009 | 16.49 |

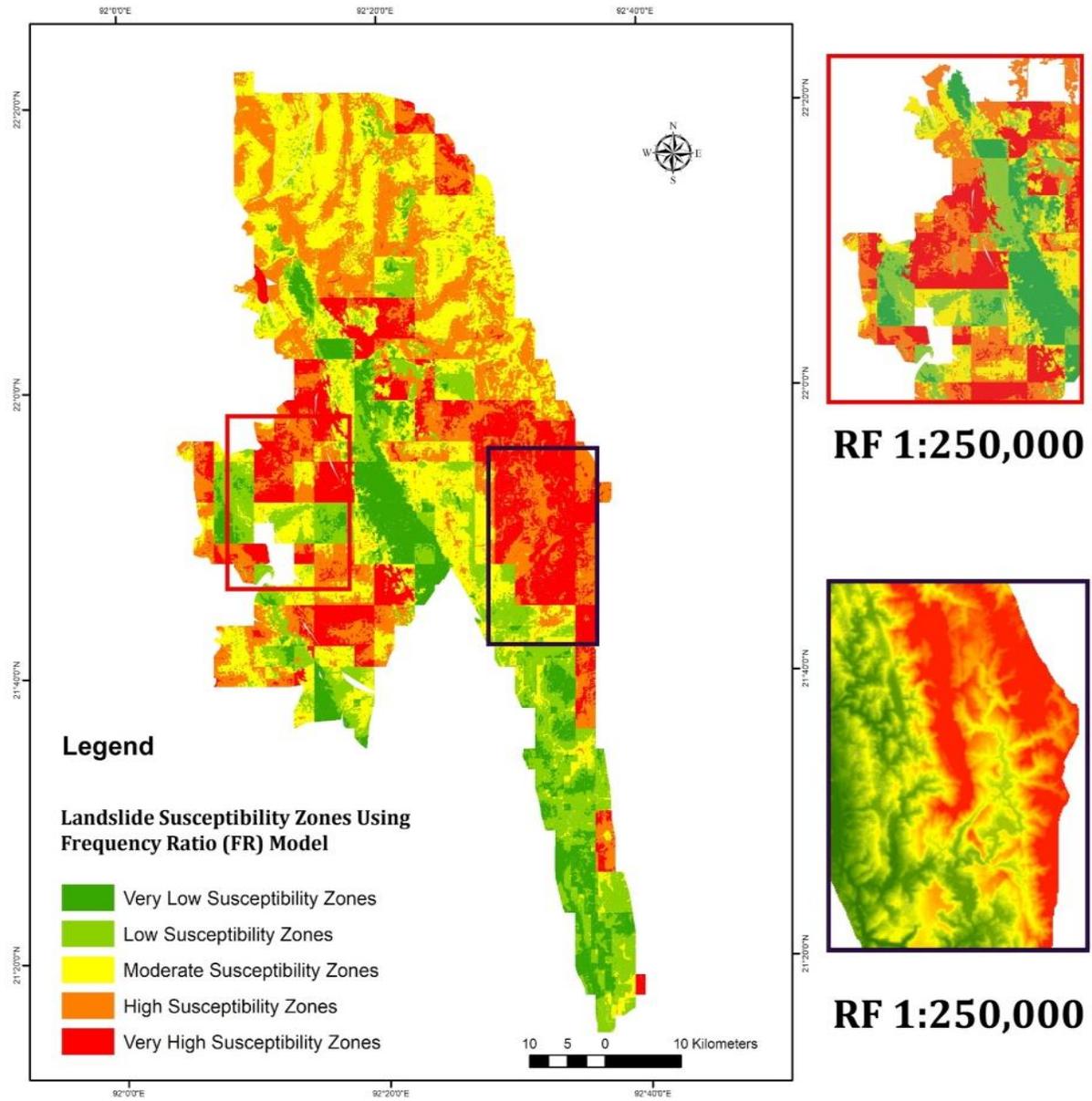

**Fig. 3:** Landslide susceptibility map of training area

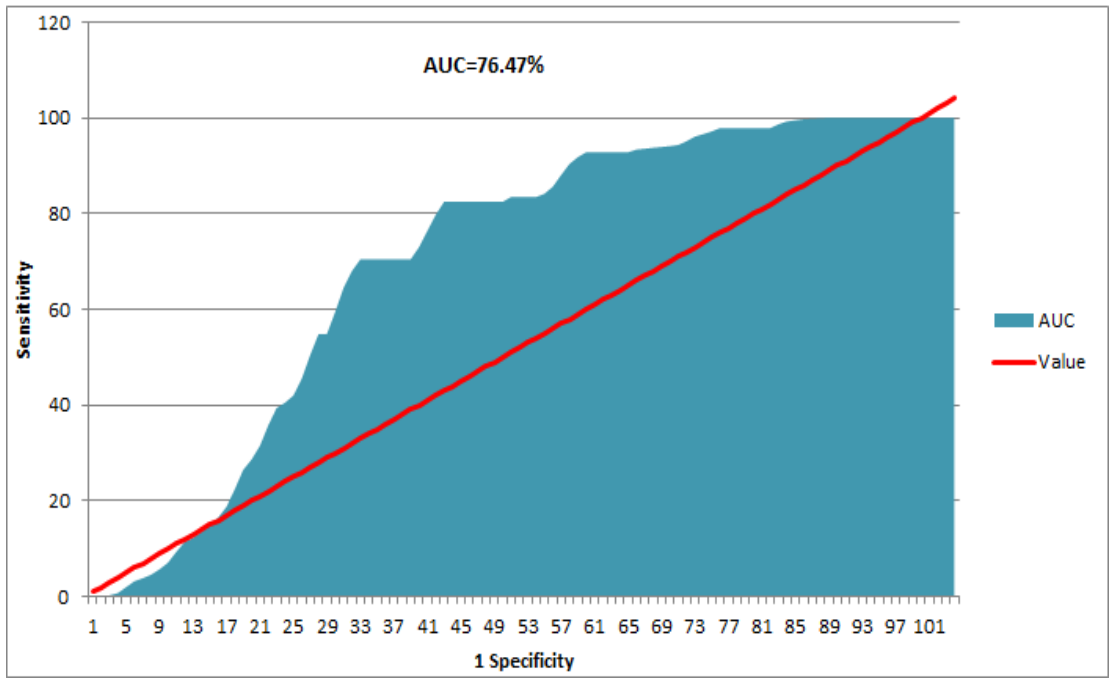

**Fig. 4.** Relative Operational Curve (ROC) of Frequency ratio model.

The results of the study include the presentation of ROC curves for landslide maps using the FR model (Figure). The calculated AUC values of the FR model were 0.7647, signifying a prediction accuracy of 76.47%. Upon analyzing the frequency ratio plot of the model (Fig. ), a noticeable trend was observed, indicating a gradual decrease in the likelihood of landslide occurrences from areas classified as very high susceptibility zones to those categorized as very low susceptibility zones. This pattern strongly suggests the presence of ideal landslide susceptibility index maps.

**Discussion**

The results of the frequency ratio (FR) model analysis in relation to landslide susceptibility in the study area reveal several noteworthy patterns and relationships between different factors and the occurrence of landslides. This assessment provides crucial insights into the dynamics of the landscape and aids in understanding the vulnerability of various areas to landslides.

The positive relationship observed between slope and landslide FR underscores the vulnerability of steep slopes to landslides. This aligns with established principles that steeper slopes are more prone to slope instability and subsequent landslides (Dai et al., 2001). Similarly, the significant role of elevation becomes evident, as higher altitude areas exhibit higher FR values, indicating a greater likelihood of landslides. This phenomenon is commonly attributed to increased rainfall and weathering at higher elevations, leading to soil saturation and reduced stability (Guzzetti et al., 1999).

The study highlights the influence of slope aspect and curvature on landslide susceptibility. Slopes facing certain directions, such as northeast-east, southeast-south, southwest-west, and northwest-north, exhibit elevated FR values, signifying an increased probability of landslides. This corresponds to established knowledge that certain aspects are more susceptible due to factors like solar radiation exposure and moisture accumulation (Van Westen et al., 2003). Additionally, the connection between curvature and landslide susceptibility aligns with geomorphic principles, where both convex and concave areas are prone to instability, often due to water accumulation and erosion processes (Montgomery and Dietrich, 1994).

The positive correlation between SPI and landslide FR emphasizes the study area's susceptibility to erosion and slope instability. High SPI values reflect erosive forces that contribute to soil erosion and enhance landslide risk (Gallant and Dowling, 2003). Moreover, the strong positive relationship between TWI and landslide frequency underscores the significance of soil saturation in landslide susceptibility.

Areas with elevated TWI values experience increased soil saturation, leading to reduced stability and heightened vulnerability to landslides (Guzzetti et al., 2005).

The findings regarding drainage density indicate that certain classes are associated with higher FR values, indicating an increased probability of landslides. This observation aligns with the understanding that high drainage density can exacerbate erosion and instability within a landscape (Pike, 2000). The study's assessment of land use changes underscores the impact of human activities on slope instability. The prevalence of land surface cutting activities and changes in landform highlights the role of anthropogenic factors in exacerbating landslide susceptibility (Aleotti and Chowdhury, 1999).

The study's identification of areas with lower Normalized Difference Vegetation Index (NDVI) values as being associated with high FR values and a greater likelihood of landslides underscores the role of vegetation cover in stabilizing slopes. Reduced vegetation cover and low density can contribute to increased soil erosion and instability, thereby enhancing landslide susceptibility (Sidle et al., 2006).

In conclusion, the results of the Frequency Ratio model analysis provide comprehensive insights into the intricate relationships between various factors and landslide susceptibility in the study area. These findings are crucial for informing effective landslide mitigation strategies and highlight the need for proactive measures to reduce the risk of landslide-related loss of lives and property.

**Conclusion**

In this study, the comprehensive analysis conducted through the FR model has unveiled critical relationships between various environmental factors and landslide. The positive correlation observed between slope and landslide FR reaffirms the vulnerability of steep slopes to landslides, aligning with established principles in geoscience. Steeper slopes, particularly those ranging from 0 to 11.44° and 27.72 to 64.61°, exhibit higher FR values, indicating a high probability of landslide occurrences. Elevation emerges as a key determinant, with higher altitude areas displaying higher FR values, attributed to increased rainfall, weathering, and soil saturation at elevated elevations.

The findings also underscore the impact of anthropogenic activities on slope instability, with significant changes identified in the study area between 1989 and 2020. Land surface cutting activities, prevalent in 53.03% of the area, highlight the role of human interventions in exacerbating landslide susceptibility. This emphasizes the need for effective land-use planning and sustainable practices to mitigate the adverse effects of such activities on slope stability. The LSM combining 13 conditioning factors facilitates a practical tool for decision-makers and land-use planners. The map classifies areas into five susceptibility zones, ranging from very high to very low, providing a nuanced understanding of the landscape's vulnerability to landslides. This information is invaluable for prioritizing monitoring efforts and implementing targeted intervention strategies in high-risk areas.

The study's presentation of ROC curves and the calculated AUC value of 0.7647 for the FR model signify a robust predictive accuracy of 76.47%. This underscores the reliability of the FR model in assessing landslide susceptibility, offering a valuable tool for early warning systems and disaster preparedness efforts. Policy implications arising from this research are crucial for informed decision-making and risk management. The identification of specific classes within factors such as drainage density, NDVI, topographic wetness index (TWI), and rainfall as high-risk contributors to landslide occurrence provides policymakers with targeted areas for intervention. Implementing land-use regulations, conservation measures, and sustainable land management practices in these identified high-risk zones can significantly reduce the likelihood and impact of landslides. Additionally, community awareness and preparedness programs should be initiated in high susceptibility zones, ensuring that residents are equipped with the knowledge and tools to respond effectively to potential landslide events. Moreover, collaboration between researchers, policymakers, and local communities can substantially contribute to successful implementation of such measures.